# Building a consistent system for faculty appraisal using Data Envelopment Analysis

Amar Oukil

*Department of Operations Management & Business Statistics,*
*College of Economics and Political Science, Sultan Qaboos University, Muscat, Oman*

aoukil@squ.edu.om

**ABSTRACT**

Data Envelopment Analysis (DEA) appears more than just an instrument of measurement. DEA models can be seen as a mathematical structure for democratic voicing within decisional contexts. Such an important aspect of DEA is enhanced through the performance evaluation of a group of professors in a virtual Business college. We show that the outcomes of the analysis can be very useful to support decision processes at many levels. There are three categories of professors: Assistant professors, Associate professors, and Full professors. The evaluation process of these professors is investigated through two different cases. The first case handles each category of professors as a separate sample representing an independent population. The results show that the mean efficiency scores fall between 0.85 and 0.93 for all professors no matters their category. In spite of enabling more fairness, such an approach suffers from its *exclusive* character, which is contrary to the *democratic* spirit of DEA. The second case tries to cope with this deficiency through the assessment of the faculty members as a single sample drawn from the same population, i.e., Assistant professors, Associate professors, and Full professors are treated equally, only on the ground of their respective inputs and outputs, no matters their academic rank. A clear efficiency decline is reported, basically due to the very nature of DEA as a procedure that is more efficiency than output focused.

## 1. Introduction

Faculty appraisal is an important process towards sustainable improvement of teaching and research quality in higher education institutions. Professors are often evaluated for different decision making purposes, like promotion, contract renewal, premium allocation, and more (see, e.g., Sun 2002, De Witte et al 2013, Sohn & Kim 2012). Whatever the outcome of such evaluations, some sort of resentment among professors is usually expected, mainly if the whole process is handled by a single person (e.g., the department chair). The sources of resentment could be: (1) the power gap between the assessor and the assessed, leading to mistrust and opposing perceptions, (2) ambiguities in the assessment procedure, which may cause doubts in the mind of the assessed regarding the assessor's accountability and *fairness*, (3) the assessed not having a *direct* "say" in his/her own appraisal (Oral *et al*. 2014). Thus, an evaluation system that would be *neutral* and *democratic* is likely to contribute immensely to dissipating such feelings. Neutrality guarantees, to some extent, more fairness, while democracy avoids exclusion and allows for genuine consideration of the assessed voice.

A broad spectrum of tools has been developed to support transparent decisions, and Data Envelopment Analysis (DEA) is one of the leading techniques.

Such an important aspect of DEA is enhanced through the performance evaluation of a group of professors in a virtual Business college. Based on different performance measures relating to teaching (student evaluations, peer ratings, advising, program development, etc.), research (publications, grants, etc.), as well as service, DEA can be used to develop an appraisal system that enables setting benchmarks for teaching/research appraisal, while supporting the assessor with decision tools that are reliable enough for guiding the assessed to the areas that may require further improvement. We show how these guidelines can be implemented for making crucial decisions in human resource management through the most standard DEA models.

The remainder of the paper is structured as follows. First, we present the DEA models selected for each stage of the study, with a focus of the practical scope of each expected outcome. Next, we move to a description of the specific context of our application, as well our perception of the problem solving approach. The last two sections are dedicated to the discussion of the results and venues for future research.

## 2. Methodological approach

The traditional procedures for evaluating professors (academic units) are based on the "output" focused approach depicted in Figure 1.

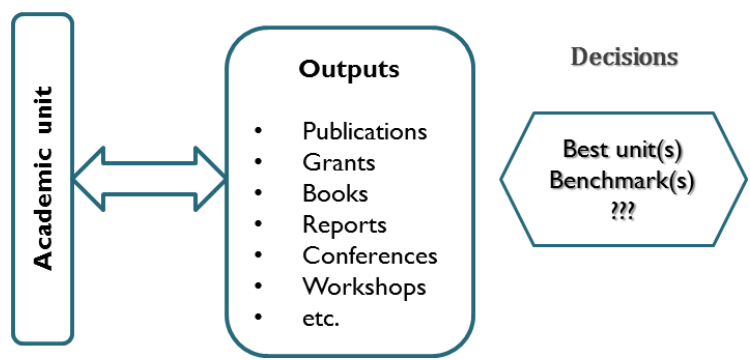

**Figure 1**. Output focused approach

As shown, the traditional procedure lies chiefly on the outputs of the candidates over a selected number of years (usually 5 years) regardless of the nature of his/her working conditions. Under a benchmarking framework, working conditions can be assimilated to the inputs of a production process (research/teaching). Indeed, although the "output" focused approach rewards the scientific contribution of the best candidate, it lacks to provide any information relating to the resources used, working conditions, skills, experience, etc. Moreover, confining the concept of best teacher/researcher to the output prevents referring to him/her as a benchmark.

A more consistent approach to the selection process must consider the professor as a production system whose output is evaluated relatively to its inputs, as displayed in Figure 2.

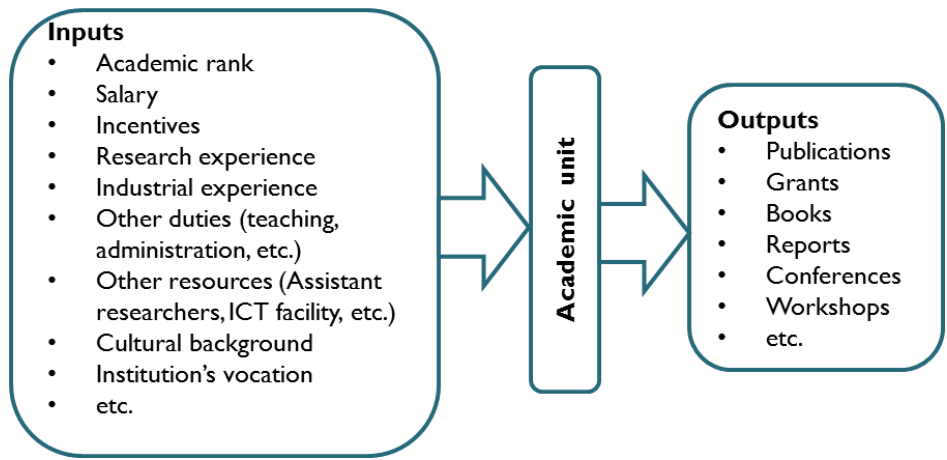

**Figure 2**. Production system approach

In the light of the proposed model, the decision spectrum becomes much broader, as depicted in Figure 3.

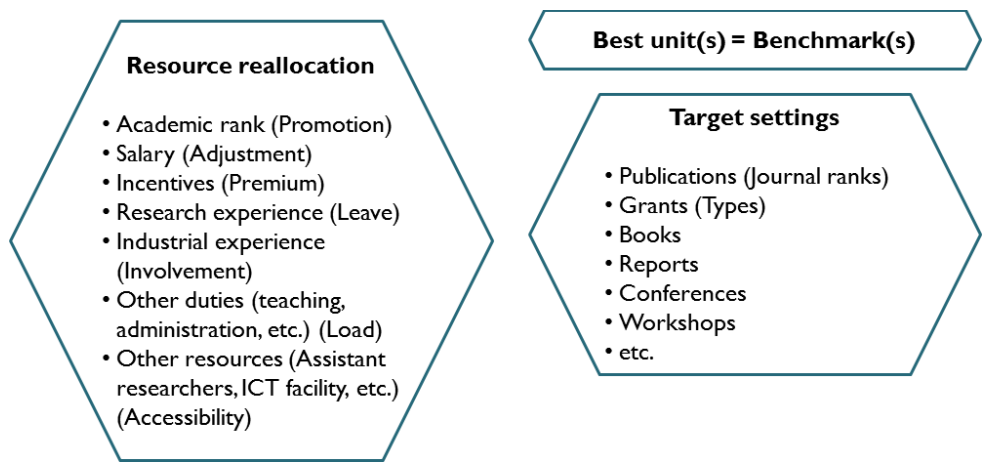

**Figure 3**. Potential decisions of the Production system approach

In order to implement the conceptual essence of the production system approach, Data Envelopment Analysis (DEA) appears the most appropriate tool for handling the multiplicity aspect of the inputs and the outputs (Sow *et al*. 2016; Oukil and Al-Zidi 2018; Al-Mezeini *et al.* 2021; Oukil 2021).

## 2.1. Model specifications

DEA is a non-parametric approach for the evaluation of relative efficiency of Decision Making Units (DMUs) conforming to an efficient production frontier. Based on mathematical programming, DEA enables not only the identification of efficiency ratios, but also the estimation of the allowable reduction of the inputs consumed by an inefficient DMU without altering any of its outputs or the required expansion of outputs produced without additional inputs (Soltani *et al.* 2021).

Regarded from a subtler angle, DEA appears more than just an instrument of measurement. DEA models can be seen as a mathematical structure for *neutral democratic voicing* within diverse decisional contexts. In the case of professor performance evaluation, the "democratic voice" is reflected via the effective participation of each decision making unit (DMU) in affirmative self-evaluation with equal rights (Oral *et al.* 2014). Interestingly, this particular feature is embodied in the self-efficiency DEA model as originally formulated by Charnes et al. (1978), CCR model henceforth. The self-evaluation DEA model not only reflects these important features but it also accentuates the concept of *appreciation* as the positive extreme of social construction (Cooperrider et al 2008, Cooperrider 2010).

*Neutrality* stems from the fact that DEA lies on a mathematical model, a structure that uses the abstract science of number, quantity, and space as a construction tool, hence, stressing the objective against the subjective and, as a result, human judgment is hypothetically excluded. Nevertheless, the exclusion of subjectivity is rather relative since human interference is present at two stages: (1) modeling, where the opinion of experts is necessary towards the best choice of the structure, the variables and parameters of the DEA model itself, (2) data collection, a task that may involve *qualitative* variables requiring quantification upon some scoring scale, which dilutes to some extent the objectivity claim.

*Democracy* is exercised via the right of every faculty member to evaluate oneself. The DEA model offers a *first-order voice* (see Oral *et al.* 2014) to each faculty in determining formally his/her own efficiency score (1) within an evaluation context shared by all faculty members (variables and parameters of the DEA model), (2) under the same evaluation conditions (constraints of the DEA model), (3) with a common evaluation criterion (objective function of the DEA model). Furthermore, the optimization process calls for the resulting mathematical model to be run for each professor individually. The first-order democratic voice is, therefore, fully expressed through such a consistent chain of formal components. Moreover, the explicit nature of the DEA model itself does highlight transparency and openness of the whole process (Oral et al. 2014).

*Appreciation* drives its meaning from the objective function of the DEA model, which defines explicitly "relative efficiency" as the core criterion of *self-evaluation*. The self-efficiency score is optimized, that is each faculty member is evaluating oneself the way he/she perceives oneself the "most favorably", hence, the appreciative feature of DEA. Such a favorable optimization confers each professor a positive appreciation at the highest level possible. In other words, the self-evaluation DEA model is the best advocate for each professor. Here again, emerges subjectivity as an accepted democratic feature that applies to every professor equally. Subsequently, Oral et al. (2014) term the self-efficiency scores as *model-based behavioral, relative, appreciative, and democratic self-evaluations*.

Indeed, the CCR model is much more than an instrument for calculating relative efficiencies as it also guarantees for each DMU the same right in evaluating themselves favorably. While the CCR model assumes constant returns to scale (CRS), the BCC model (Banker et al. 1984) is its counterpart that allows variable returns to scale (VRS). These models are formulated as linear programs (LPs) as described in what follows.

## 2.2. Standard DEA models

Assume a set of $K$ professors, each professor $k$ defined with $N$ inputs $x$ and $M$ outputs $y$. With reference to the underlying production technology, professor $(x_k, y_k)$ is fully defined with the observed values $x_{ik}$ and $y_{jk}$, $i=1,..., N$ and $j=1,..., M$. In order to estimate the efficiency score $\theta$ of professor $(x_0, y_0)$ and set production targets for inefficient professors, the input-oriented formulation of CCR model can be represented as follows (Amin and Oukil, 2019b).

$$(\text{CCR}) \quad \min \theta \qquad (1)$$

$$s.t. \quad \sum_{k=1}^{K} \lambda_k x_{ik} \leq \theta x_{i0} \quad i = 1, \dots, N \qquad (2)$$

$$\sum_{k=1}^{K} \lambda_k y_{jk} \geq y_{j0} \quad j = 1, \dots, M \qquad (3)$$

$$\lambda_k \geq 0 \quad k = 1, \dots, K \qquad (4)$$

The efficiency $\theta$ of professor $(x_0, y_0)$ is assessed by calculating the minimal radial reduction of inputs that is required to reach the efficiency frontier for a specified level of outputs. $\lambda$ measures the weights of the peers in producing the projection of professor $(x_0, y_0)$ on the efficiency frontier. Constraints (2) and (3) state that reference points are linear combinations of the input and output values of efficient peers for professor $(x_0, y_0)$.

(CCR) represents a LP model with $N+M$ constraints (not counting the non-negativity constraints) and must be solved $K$ times, once for each professor. BCC model can be obtained from (CCR) by adding the convexity constraint that guarantees that only weighted averages of efficient professors enter the reference set, i.e.

$$\sum_{k=1}^{K} \lambda_k = 1 \qquad (5)$$

CCR and BCC models are both formulated with the implicit assumption that the assessed professors operate within homogeneous environments, which presupposes that only variables representing proper inputs are integral part of the production technology.

## 2.3. Scale efficiency

Let $\theta^*_{CRS}$ and $\theta^*_{VRS}$ denote the aggregate and technical efficiency scores of professor ($x_0$, $y_0$) calculated using CRS and VRS models, respectively.

The scale efficiency *SE* of a professor is the ratio of the aggregate efficiency $\theta^*_{CRS}$ over the technical efficiency $\theta^*_{VRS}$. The values of SE can be either one or less than one.

A professor is scale efficient when its scale efficiency is equal to one, suggesting that the professor is operating at the most productive scale size and any alteration on its size will lead to inefficiency. Scale inefficiency occurs for values of SE less than one, due to either increasing or decreasing returns to scale. Following Banker et al. (2004), if $\lambda^*$ is an optimal solution of CCR model and $\sum_{k=1}^{K} \lambda_k^* > 1$, we can say that the professor exhibits decreasing return to scale (DRS), implying that the professor is operating at a scale greater than the most productive scale size of the inputs. Conversely, $\sum_{k=1}^{K} \lambda_k^* < 1$ suggests that the professor is operating in the increasing returns to scale (IRS) region, at a scale smaller than the most productive scale. The managerial interpretation of the latter inference is that the average productivity can be increased if the level of outputs increases as a result of a proportional increase in the consumption of the inputs (Banker and Morey, 1986). This can be achieved by transferring resources from professors operating at DRS to those operating at IRS (Boussofiane et al. (1992). Constant returns to scale, i.e., $\sum_{k=1}^{K} \lambda_k^* = 1$, imply that the professor is scale efficient.

## 3. Application context and procedures

The DEA methodology has been developed within the same context described in Oral et al. (2014). Five output variables are considered, namely: the annual average number of peer reviewed articles published during the last 5 years in the academic journals recognized by the business school, the annual average number of peer reviewed articles published in proceedings during the last 5 years, the annual average monetary contributions to the school through research funding and consulting projects during the last 5 years, the teaching scores of the faculty members which includes the composite scores reflecting evaluations of both the students and the Dean, and the citizenship score of each faculty member that measures the involvement in committee works, holding administrative positions, contributions to the reputation of the school. The only input variable of the model is the salary of each professor. Teaching and citizenship scores are on a scale of 0 (minimum) and 5 (maximum) whereas the other scores are actual numbers.

Only 32 faculty members are considered in Oral *et al.* (2014), including 9 full professors, 12 associate professors, and 11 assistant professors. In order to run the DEA models separately for each group of professors, a minimum number $\gamma$ of observations (professors) is required for a clear efficiency discrimination within each group. Cooper et al. (2002, Chapter 9, p.252) suggest that $\gamma$ must be greater than or equal to max{$mn$, $3(m+n)$}, where *m* is the number of outputs and *n* the number of inputs. Following the latter, we need to have at least 18 professors in each group. Accordingly, we used Monte-Carlo variate generation to expand each group with 20 "virtual" professors. Group sizes of 29 full professors, 32 associate professors, and 31 assistant professors are comfortably large to run the DEA models for each group of professors separately, using the variables summarized in Table 1. The full data set is available from the author.

The evaluation process of these professors is conducted over two scenarios. Firstly, each category of professors is handled as a separate sample representing a different population. Under the assumption of CRS, such an approach enables fairness but suffers from its exclusive character, which is contrary to the democratic spirit of DEA.

**Table 1**. Summary statistics of variables in efficiency measurement

| | Variables | OUTPUTS | | | | | INPUT |
|---|---|---|---|---|---|---|---|
| | | Peer Reviewed papers (per year) | Other publications & proceedings (per year) | Total Research funding ($ 000) | Quality of teaching | Administrative involvement | Present Annual salary ($ 000) |
| Full professors | Mean | 0.55 | 0.64 | 89.28 | 3.54 | 2.57 | 174.04 |
| | STD | 0.31 | 0.30 | 49.63 | 0.94 | 1.29 | 34.98 |
| | Min | 0.10 | 0.15 | 12.00 | 2.20 | 0.34 | 103.18 |
| | Max | 1.12 | 1.29 | 176.00 | 5.00 | 4.76 | 277.88 |
| Associate professors | Mean | 0.60 | 0.83 | 60.23 | 3.53 | 3.33 | 132.25 |
| | STD | 0.28 | 0.49 | 38.50 | 0.61 | 0.78 | 8.26 |
| | Min | 0.10 | 0.10 | 1.21 | 2.50 | 2.10 | 113.43 |
| | Max | 1.00 | 1.60 | 136.63 | 4.60 | 4.60 | 147.55 |
| Assistant professors | Mean | 0.24 | 0.25 | 15.04 | 3.56 | 1.78 | 89.05 |
| | STD | 0.20 | 0.20 | 8.85 | 0.68 | 1.40 | 12.64 |
| | Min | 0.00 | 0.00 | 1.02 | 2.07 | 0.00 | 59.62 |
| | Max | 0.60 | 0.60 | 34.00 | 5.00 | 4.10 | 113.95 |

Hence, the second scenario, the faculty members are merged as a single sample drawn from the same population, i.e., Assistant professors, Associate professors, and Full professors are treated equally, only on the ground of their respective inputs and outputs, no matters their academic rank. Under the assumption of CRS, this approach may lead to conclusions that might not be too realistic as one of the core concepts of DEA, fairness, is violated. The remedy to these deficiencies is probably the usage of the latter merged sample under VRS. The results and conclusions are developed in the next section.

## 4. Results and discussion

The DEA models pertaining to the case study described in Section 3 have been implemented and solved using a C++ code, embedding IBM-ILOG CPLEX version 12.4. The code computes the optimal efficiency scores $\theta^*_{CRS}$ and $\theta^*_{VRS}$ for each professor, besides the corresponding optimal solution $\lambda^*$ and the slack values. We consider the two scenarios separately. separate faculty categories and a merged faculty sample.

### 4.1. Separate faculty categories

As explained in Section 3, this case considers the evaluation of the faculty members within their own rank categories. In other words, a full professor is evaluated relatively to other full professors, and so forth.

### a) Efficiency scores

As displayed in Table 2, the average efficiency scores of full professors, associate professors, and assistant professors are 0.85, 0.93 and 0.87, respectively. This suggests that salaries could be reduced by 15%, 7% and 13%, on average, for each professor's category, respectively, and the professor will still produce the same level of output.

On another hand, 6 full professors out of 29 (20.69%), 15 associate professors out of 32 (46.87%), and 6 assistant professors out of 31 (19.35%) are found to be efficient, meaning that they are, relatively, using the input (salary) the best way possible.

Over the three professors' categories involved in the study, 15 over 27 (55.55%) efficient professors are located in the associate professors group, with the highest mean efficiency score of 0.93. Only 1 associate professor showed a performance below 0.70 whilst 25 professors (almost 80%) scored an efficiency of more than 0.90. Meanwhile, the efficiency trends are just about identical for full and assistant professors' categories. In each of these categories, the efficiency score is less than 0.70 for 5 professors and it exceeds 0.90 for 14 professors (almost 50%).

**Table 2**. Frequency distributions of the efficiency scores $\theta^*_{CRS}$

| Efficiency scores | Full professors | Associate professors | Assistant professors |
|---|---|---|---|
| < 0.5 | 1 | 0 | 0 |
| 0.5-0.6 | 1 | 0 | 0 |
| 0.6-0.7 | 3 | 1 | 5 |
| 0.7-0.8 | 5 | 3 | 5 |
| 0.8-0.9 | 5 | 3 | 7 |
| 0.9-1.0 | 8 | 10 | 8 |
| 1.0 | 6 | 15 | 6 |
| **Mean** | 0.85 | 0.93 | 0.87 |
| **STD** | 0.15 | 0.10 | 0.12 |
| **Min** | 0.43 | 0.67 | 0.65 |

## b) Scale effects and Returns to scale

In order to investigate the effect of salary level on the faculty performance, we calculated the scale efficiency for each professor as well as the corresponding returns to scale using the formulae presented in Section 2.2. Table 3 provides the mean scale efficiency SE for the professors besides their characteristics with respect to returns to scale. The results show a mean scale efficiency of more than 0.93 for the professors, whatever their category, indicating that the majority of professors are operating near the optimal salary level, with more than two thirds of the professors exceeding the scale level of 0.95 (Table 1A to 3A in the appendices). Hence, we can say that scale efficiencies are relatively high, suggesting that inefficiency is mainly due to inadequate use of input (salary) rather than to the input size (salary level). With respect to this aspect, a total of 23 professors are experiencing decreasing returns to scale, which suggests that a quarter of the sampled professors are not using their resources (salary) properly, hence the moderate level of inefficiency. In other words, the productivity of these professors can, on average, be increased through a proportional increase in the consumption of inputs. In the context of salaries as the only input, the process is far from being an ordinary transformation of inputs into outputs, making the latter inference not practical. Instead, salary readjustment can be envisaged through a reassignment of resources from sub-optimal to super-optimal professors (salary compression/inversion) on a case by case basis using the efficiency score of each professor. The professors who fall under the super-optimal class are those exhibiting increasing returns to scale, and we count 40 such professors out of 92.

**Table 3**. Characteristics of salaries (in $000) with respect to returns to scale

| Category | SE | Super Optimal | | Optimal | | Sub-optimal | |
|---|---|---|---|---|---|---|---|
| | | # professors | Mean | # professors | Mean | # professors | Mean |
| Full | 0.93 | 11 | 167 | 6 | 143 | 12 | 196 |
| Associate | 0.96 | 10 | 128 | 17 | 132 | 5 | 141 |
| Assistant | 0.94 | 19 | 85 | 6 | 90 | 6 | 101 |
| Total | | 40 | | 29 | | 23 | |

**Table 4**. Input slacks and excess inputs

| Category | Mean slack | Mean input use | Excess input (%) |
|---|---|---|---|
| **Full** | 34.86 | 182.04 | 19.15 |
| **Associate** | 15.96 | 132.66 | 12.03 |
| **Assistant** | 14.54 | 88.93 | 16.35 |

In order to estimate the excess of input that can be transferred or, eventually deducted, we calculate the slack variables corresponding to the resource salary for each professor. In theory, the slack values provide flexible margins for salary reduction without altering the existing outputs. The mean slack values are summarized in Table 4, together with the average usage of inputs and the corresponding proportions of excess input usage. The largest excess of input use is detected for the salaries of full professors. Actually, the salaries of 12 full professors might be reduced by 19.15% on average while maintaining their production level unchanged.

On another hand, average reductions of 12.03% and 16.03% can also be applied on the salaries of 5 associate and 6 assistant professors, respectively.

Out of 92 professors, 28 showed constant returns to scale. These professors belong to the optimal class, which is used as a benchmark. Table 5 summarizes the most relevant features of the optimal professors from each category.

**Table 5**. Characteristics of the Optimal professors

| | Variables | OUTPUTS | | | | | INPUT |
|---|---|---|---|---|---|---|---|
| | | Peer Reviewed papers (per year) | Other publications & proceedings (per year) | Total Research funding ($ 000) | Quality of teaching | Administrative involvement | Present Annual salary ($ 000) |
| Full professor | Mean | 0.60 | 0.79 | 101.03 | 3.47 | 2.34 | 143 |
| | Max | 0.94 | 1.29 | 148.00 | 4.90 | 4.80 | 173 |
| | Min | 0.19 | 0.24 | 30.00 | 2.40 | 0.60 | 103 |
| Associate professor | Mean | 0.64 | 0.86 | 61.31 | 3.61 | 3.48 | 132 |
| | Max | 1.00 | 1.60 | 137.00 | 4.60 | 148.00 | 148 |
| | Min | 0.10 | 0.10 | 3.00 | 2.50 | 118.00 | 118 |
| Assistant professor | Mean | 0.40 | 0.39 | 23.23 | 4.20 | 2.27 | 90 |
| | Max | 0.58 | 0.60 | 31.00 | 5.30 | 4.10 | 102 |
| | Max | 0.12 | 0.04 | 17.00 | 3.50 | 0.10 | 75 |

Primarily, the expected levels for optimal salaries are $ 143K/year, $ 132K/year, and $ 87K/year for full professors, associate professors and assistant professors, respectively.

### 4.2. Merged faculty sample

In this section, the evaluation process is investigated throughout a single sample, obtained by merging all professors into the same group. Such merger assumes implicitly that all professors, whatever their ranks, are subjected to equal opportunities, enjoy the same advantages, and possess identical potential. In what follows, we will show that these assumptions are unrealistic and, obviously, lead to a biased decision scheme. For the sake of comparison, the results will be displayed according to the same format used in Section 4.1.

### a) Efficiency scores

Table 6 reveals that the average efficiency scores of full professors, associate professors, and assistant professors are 0.72, 0.88 and 0.85, respectively. It is clear that there is a deterioration of the performance indices, compared to the "separate faculty categories" case. The deterioration is more acute for the full professors class, with an average dropping gap of 13% but, less severe for assistant professors (-2%). Such a trend becomes more tangible if we note that the number of efficient professors decreased from 6 to 1 for full professors and from 15 to 10 for associate professors. Moreover, the efficiency has declined for all professors, no matters their rank. Interestingly, the frequency distribution of the efficiency scores is more or less preserved for assistant professors. Therefore, the standard DEA model appears more in favour of assistant professors rather than full professors. In other words, DMUs with higher inputs (salary) are under-rated.

**Table 6**. Frequency distributions of the efficiency scores $\theta^*_{CRS}$

| Efficiency scores | Full professors | Associate professors | Assistant professors |
|---|---|---|---|
| < 0.5 | 2 | 0 | 0 |
| 0.5-0.6 | 3 | 0 | 0 |
| 0.6-0.7 | 8 | 6 | 6 |
| 0.7-0.8 | 9 | 4 | 5 |
| 0.8-0.9 | 3 | 4 | 8 |
| 0.9-1.0 | 3 | 8 | 6 |
| 1.0 | 1 | 10 | 6 |
| **Mean** | 0.72 | 0.88 | 0.85 |
| **STD** | 0.15 | 0.13 | 0.12 |
| **Min** | 0.37 | 0.65 | 0.65 |

The plausible reason of such discrepancy resides in the very nature of DEA as a model that is more efficiency than output concerned, i.e., it gives much more priority to the salary and how well it is used.

**b) Scale effects and Returns to scale**
Following the same reasoning as in Section 3, the results in Table 7 show a mean scale efficiency exceeding 0.91 for all professors, regardless their rank, with 67.4% of these values over 0.95 (Tables 1B to 3B in the appendices). This not only confirms that the salaries of most professors are close to the optimal salary level but also that inefficiency is caused by inadequate use of salary rather than its size. In addition, this supports the choice of CCR model as an appropriate DEA tool for the evaluation. The number of professors experiencing decreasing returns to scale increased from 23 to 29, which explains the efficiency decline. The optimal and super-optimal classes count 17 and 46 professors, respectively, that is 11 optimal professors less than the previous case.
The slack information corresponding to the present case is given in Table 8. Once again, the salaries of full professors exhibit the largest excess of input use, nearing on average 30.12%.

**Table 7.** Characteristics of salaries (in $000) with respect to returns to scale

| Category | SE | Super Optimal | | Optimal | | Sub-optimal | |
|---|---|---|---|---|---|---|---|
| | | # professors | Mean | # professors | Mean | # professors | Mean |
| Full | 0.91 | 11 | 157 | 1 | 155 | 17 | 186 |
| Associate | 0.96 | 12 | 129 | 10 | 130 | 10 | 139 |
| Assistant | 0.94 | 23 | 87 | 6 | 90 | 2 | 113 |
| Merger | 0.94 | 46 | 115 | 17 | 117 | 29 | 165 |

An average excess of 17.28% and 18.72% is also indicated on the salaries of associate and assistant professors, respectively.

**Table 8.** Input slacks and excess inputs

| Category | Mean slack | Mean input use | Excess input (%) |
|---|---|---|---|
| **Full** | 52.62 | 174.73 | 30.12 |
| **Associate** | 23.00 | 133.14 | 17.28 |
| **Assistant** | 16.65 | 88.93 | 18.72 |

Table 9 sums up the most important information regarding the optimal professors by rank. Most importantly, the expected levels for optimal salaries are $ 155K/year, $ 130K/year, and $ 90K/year for full professors, associate professors and assistant professors, respectively. These figures show more appreciation to the salaries of full professors in comparison with the previous case.

**Table 9.** Characteristics of the Optimal professors

| | Variables | OUTPUTS | | | | | INPUT |
|---|---|---|---|---|---|---|---|
| | | Peer Reviewed papers (per year) | Other publications & proceedings (per year) | Total Research funding ($ 000) | Quality of teaching | Administrative involvement | Present Annual salary ($ 000) |
| **Full professor** | Mean | 0.67 | 0.94 | 147.76 | 2.37 | 4.76 | 155 |
| | Max | 0.67 | 0.94 | 147.76 | 2.37 | 4.76 | 155 |
| | Min | 0.67 | 0.94 | 147.76 | 2.37 | 4.76 | 155 |
| **Associate professor** | Mean | 0.72 | 1.14 | 81.38 | 3.61 | 3.21 | 130 |
| | Max | 1.00 | 1.60 | 137.00 | 4.40 | 4.50 | 148 |
| | Min | 0.30 | 0.30 | 29.00 | 2.50 | 2.10 | 118 |
| **Assistant professor** | Mean | 0.40 | 0.39 | 23.23 | 4.20 | 2.27 | 90 |
| | Max | 0.58 | 0.60 | 31.00 | 5.30 | 4.10 | 102 |
| | Max | 0.12 | 0.04 | 17.00 | 3.50 | 0.10 | 75 |

## 5. Conclusions & recommendations

Our study was concerned with evaluating the performance of professors from a virtual college of business. Out of 92 professors, we used real life data for 32 professors and we generated random data for the remaining 60. We considered two cases: the "merged faculty sample" and the "separate faculty categories". The results of the efficiency scores reveal that there is more discrimination among professors of the "merged faculty sample" case than those of the "separate faculty categories" case. This situation can be imputed to the stability property of DEA, establishing that the more DMUs the more discrimination. Moreover, the merger discards the professor rank in spite of its importance in defining the statistical structure of the data itself. Such an aspect can be handled within a hierarchical DEA model as a contextual variable.

The scale efficiency scores of more than two thirds the number of professors is higher than 0.95 in both cases. This is an indication strong enough to assert that using DEA under constant returns to scale is the right approach to investigate the efficiency of professors, irrespective their ranks.

The excess salary ratios over the two cases fall between 12% and 30%, which may primarily reflect an overestimation of the salaries. This statement, even important, cannot imply "immediate" actions, like salary compression but, it may help in readjusting salaries offered to new faculty members. The latter decision can be supported through the salary benchmarks calculated for the optimal professors. The mean values of inputs and outputs, as illustrated in Tables 5 and 9 can also be employed to set reliable *standards* for the appraisal of professors working within the same academic structure (department, college, university) under similar conditions. In addition, they can be used to shape some benchmark to assist promotion committees.

Although efficiency scores enable the identification of the most efficient professors, i.e. those who use the resource salary the best but they cannot be used directly to fully rank professors because of the multiple occurrences of "1" as an efficiency score. As an alternative, future research may consider using cross-efficiency DEA models jointly with some aggregation model to achieve more discrimination among the assessed professors; See, e.g., Oral *et al*. (2015), Oukil and Amin (2015), Amin and Oukil, (2019a), Oukil (2020a, 2020b), Oukil and Govindaluri (2020), Oukil and El-Bouri (2021).

Another venue of research may consist in investigating the impact of contextual factors relating to the academia on the performance of the faculty. These aspects can be handled through two stage frameworks that involve DEA and some econometric models (Oukil *et al*, 2016; Oukil and Zekri 2014, 2021).

# APPENDICES

Table 1A. Specific efficiency scores of Full professors

| Professor | $\theta^*_{CRS}$ | $\theta^*_{VRS}$ | SE | $\sum_{k=1}^{K} \lambda^*_k$ | Status |
|---|---|---|---|---|---|
| 1 | 0.94 | 1.00 | 0.94 | 1.52 | *decr.* |
| 2 | 0.88 | 0.91 | 0.96 | 1.25 | *decr.* |
| 3 | 0.89 | 0.92 | 0.97 | 1.19 | *decr.* |
| 4 | 0.87 | 0.88 | 0.99 | 1.02 | *decr.* |
| 5 | 0.70 | 0.83 | 0.84 | 0.77 | *incr.* |
| 6 | 0.78 | 0.82 | 0.96 | 0.90 | *incr.* |
| 7 | 0.80 | 0.89 | 0.90 | 0.84 | *incr.* |
| 8 | 1.00 | 1.00 | 1.00 | 1.00 | *const.* |
| 9 | 0.91 | 1.00 | 0.91 | 1.30 | *decr.* |
| 10 | 0.43 | 0.91 | 0.47 | 0.45 | *incr.* |
| 11 | 1.00 | 1.00 | 1.00 | 1.00 | *const.* |
| 12 | 0.74 | 0.85 | 0.87 | 0.82 | *incr.* |
| 13 | 0.94 | 0.97 | 0.98 | 1.06 | *decr.* |
| 14 | 0.96 | 0.99 | 0.97 | 1.10 | *decr.* |
| 15 | 0.83 | 0.84 | 0.99 | 0.98 | *incr.* |
| 16 | 0.73 | 0.74 | 0.99 | 0.94 | *incr.* |
| 17 | 0.63 | 1.00 | 0.63 | 1.17 | *decr.* |
| 18 | 1.00 | 1.00 | 1.00 | 1.00 | *const.* |
| 19 | 0.94 | 1.00 | 0.94 | 1.09 | *decr.* |
| 20 | 0.72 | 0.74 | 0.98 | 0.96 | *incr.* |
| 21 | 0.91 | 1.00 | 0.91 | 1.35 | *decr.* |
| 22 | 0.85 | 0.87 | 0.97 | 0.95 | *incr.* |
| 23 | 0.55 | 0.62 | 0.88 | 0.84 | *incr.* |
| 24 | 1.00 | 1.00 | 1.00 | 1.00 | *const.* |
| 25 | 0.91 | 0.96 | 0.95 | 1.44 | *decr.* |
| 26 | 1.00 | 1.00 | 1.00 | 1.00 | *const.* |
| 27 | 1.00 | 1.00 | 1.00 | 1.00 | *const.* |
| 28 | 0.96 | 1.00 | 0.96 | 1.36 | *decr.* |
| 29 | 0.63 | 0.73 | 0.86 | 0.78 | *incr.* |
| **Average** | 0.85 | 0.91 | 0.93 | | |

Table 2A. Specific efficiency scores of Associate professors

| Professor | $\theta^*_{CRS}$ | $\theta^*_{VRS}$ | SE | $\sum_{k=1}^{K} \lambda^*_k$ | Status |
|---|---|---|---|---|---|
| 1 | 1.00 | 1.00 | 1.00 | 1.00 | *const.* |
| 2 | 1.00 | 1.00 | 1.00 | 1.00 | *const.* |
| 3 | 0.89 | 0.93 | 0.95 | 0.83 | *incr.* |
| 4 | 0.91 | 0.91 | 1.00 | 1.00 | *const.* |
| 5 | 0.91 | 0.92 | 0.99 | 1.02 | *decr.* |
| 6 | 0.99 | 1.00 | 0.99 | 1.09 | *decr.* |
| 7 | 0.90 | 0.90 | 1.00 | 1.00 | *const.* |
| 8 | 0.95 | 1.00 | 0.95 | 1.12 | *decr.* |
| 9 | 1.00 | 1.00 | 1.00 | 1.10 | *decr.* |
| 10 | 1.00 | 1.00 | 1.00 | 1.00 | *const.* |
| 11 | 1.00 | 1.00 | 1.00 | 1.00 | *const.* |
| 12 | 1.00 | 1.00 | 1.00 | 1.00 | *const.* |
| 13 | 0.67 | 1.00 | 0.67 | 0.63 | *incr.* |
| 14 | 1.00 | 1.00 | 1.00 | 1.00 | *const.* |
| 15 | 0.90 | 0.90 | 1.00 | 0.95 | *incr.* |
| 16 | 0.95 | 0.97 | 0.98 | 0.90 | *incr.* |
| 17 | 1.00 | 1.00 | 1.00 | 1.00 | *const.* |
| 18 | 1.00 | 1.00 | 1.00 | 1.00 | *const.* |
| 19 | 0.91 | 0.95 | 0.96 | 0.84 | *incr.* |
| 20 | 0.71 | 0.97 | 0.73 | 0.65 | *incr.* |
| 21 | 0.99 | 1.00 | 0.99 | 0.91 | *incr.* |
| 22 | 1.00 | 1.00 | 1.00 | 1.00 | *const.* |
| 23 | 1.00 | 1.00 | 1.00 | 1.00 | *const.* |
| 24 | 1.00 | 1.00 | 1.00 | 1.00 | *const.* |
| 25 | 1.00 | 1.00 | 1.00 | 1.00 | *const.* |
| 26 | 0.85 | 0.86 | 0.99 | 1.02 | *decr.* |
| 27 | 0.72 | 0.85 | 0.85 | 0.83 | *incr.* |
| 28 | 1.00 | 1.00 | 1.00 | 1.00 | *const.* |
| 29 | 0.75 | 0.92 | 0.82 | 0.75 | *incr.* |
| 30 | 0.93 | 1.00 | 0.93 | 0.90 | *incr.* |
| 31 | 1.00 | 1.00 | 1.00 | 1.00 | *const.* |
| 32 | 1.00 | 1.00 | 1.00 | 1.00 | *const.* |
| **Average** | 0.93 | 0.97 | 0.96 | | |

Table 3A. Specific efficiency scores of Assistant professors

| Professor | $\theta^*_{CRS}$ | $\theta^*_{VRS}$ | SE | $\sum_{k=1}^{K} \lambda_k^*$ | Status |
|---|---|---|---|---|---|
| 1 | 0.98 | 1.00 | 0.98 | 1.20 | decr. |
| 2 | 0.95 | 0.95 | 1.00 | 0.98 | incr. |
| 3 | 0.76 | 0.79 | 0.96 | 0.84 | incr. |
| 4 | 0.98 | 1.00 | 0.98 | 1.02 | decr. |
| 5 | 0.83 | 0.87 | 0.96 | 0.92 | incr. |
| 6 | 1.00 | 1.00 | 1.00 | 1.00 | const. |
| 7 | 0.89 | 0.90 | 0.99 | 0.90 | incr. |
| 8 | 0.92 | 0.93 | 0.99 | 0.94 | incr. |
| 9 | 0.81 | 0.93 | 0.87 | 0.69 | incr. |
| 10 | 0.79 | 0.93 | 0.85 | 0.62 | incr. |
| 11 | 0.88 | 1.00 | 0.88 | 0.68 | incr. |
| 12 | 0.66 | 1.00 | 0.66 | 0.43 | incr. |
| 13 | 0.65 | 0.73 | 0.89 | 0.70 | incr. |
| 14 | 0.96 | 0.97 | 0.98 | 1.06 | decr. |
| 15 | 0.95 | 0.96 | 0.98 | 0.88 | incr. |
| 16 | 0.85 | 0.88 | 0.97 | 1.03 | decr. |
| 17 | 1.00 | 1.00 | 1.00 | 1.00 | const. |
| 18 | 1.00 | 1.00 | 1.00 | 1.00 | const. |
| 19 | 0.94 | 1.00 | 0.94 | 0.65 | incr. |
| 20 | 0.88 | 0.88 | 0.99 | 0.98 | incr. |
| 21 | 0.87 | 0.87 | 1.00 | 1.02 | decr. |
| 22 | 0.67 | 0.83 | 0.80 | 0.55 | incr. |
| 23 | 1.00 | 1.00 | 1.00 | 1.00 | incr. |
| 24 | 1.00 | 1.00 | 1.00 | 1.00 | const. |
| 25 | 0.76 | 0.81 | 0.95 | 0.78 | incr. |
| 26 | 1.00 | 1.00 | 1.00 | 1.00 | const. |
| 27 | 0.68 | 0.84 | 0.82 | 0.65 | incr. |
| 28 | 0.91 | 1.00 | 0.91 | 0.84 | incr. |
| 29 | 0.79 | 0.81 | 0.98 | 1.02 | decr. |
| 30 | 0.69 | 0.75 | 0.93 | 0.74 | incr. |
| 31 | 0.80 | 0.83 | 0.97 | 0.78 | incr. |
| Average | 0.87 | 0.92 | 0.94 | | |

Table 1B. Specific efficiency scores of Full professors

| Professor | $\theta^*_{CRS}$ | $\theta^*_{VRS}$ | SE | $\sum_{k=1}^{K} \lambda_k^*$ | Status |
|---|---|---|---|---|---|
| 1 | 0.82 | 1.00 | 0.82 | 1.41 | decr. |
| 2 | 0.74 | 0.82 | 0.90 | 1.13 | decr. |
| 3 | 0.77 | 0.82 | 0.94 | 1.15 | decr. |
| 4 | 0.66 | 0.66 | 0.99 | 0.97 | incr. |
| 5 | 0.64 | 0.65 | 0.99 | 0.94 | incr. |
| 6 | 0.61 | 0.62 | 0.99 | 0.90 | incr. |
| 7 | 0.70 | 0.73 | 0.95 | 0.86 | incr. |
| 8 | 0.85 | 0.86 | 0.99 | 1.03 | decr. |
| 9 | 0.70 | 1.00 | 0.70 | 1.18 | decr. |
| 10 | 0.37 | 0.53 | 0.69 | 0.44 | incr. |
| 11 | 1.00 | 1.00 | 1.00 | 1.00 | const. |
| 12 | 0.65 | 0.66 | 0.98 | 0.90 | incr. |
| 13 | 0.70 | 0.70 | 1.00 | 1.00 | decr. |
| 14 | 0.80 | 0.82 | 0.97 | 1.03 | decr. |
| 15 | 0.62 | 0.63 | 0.99 | 0.92 | incr. |
| 16 | 0.62 | 0.64 | 0.98 | 0.92 | incr. |
| 17 | 0.52 | 1.00 | 0.52 | 1.14 | decr. |
| 18 | 0.98 | 1.00 | 0.98 | 1.21 | decr. |
| 19 | 0.77 | 0.77 | 1.00 | 1.01 | decr. |
| 20 | 0.50 | 0.66 | 0.75 | 1.15 | decr. |
| 21 | 0.81 | 1.00 | 0.81 | 1.35 | decr. |
| 22 | 0.69 | 0.71 | 0.97 | 0.92 | incr. |
| 23 | 0.51 | 0.51 | 1.00 | 1.01 | decr. |
| 24 | 0.96 | 0.99 | 0.97 | 1.07 | decr. |
| 25 | 0.76 | 0.95 | 0.79 | 1.27 | decr. |
| 26 | 0.90 | 0.92 | 0.98 | 0.95 | incr. |
| 27 | 0.79 | 0.80 | 0.99 | 1.03 | decr. |
| 28 | 0.79 | 1.00 | 0.79 | 1.25 | decr. |
| 29 | 0.53 | 0.60 | 0.88 | 0.71 | incr. |
| Average | 0.72 | 0.80 | 0.91 | | |

**Table 2B**. Specific efficiency scores of Associate professors

| Professor | $\theta^*_{CRS}$ | $\theta^*_{VRS}$ | SE | $\sum_{k=1}^{K} \lambda^*_k$ | Status |
|---|---|---|---|---|---|
| 1 | 0.72 | 1.00 | 0.72 | 1.17 | *decr.* |
| 2 | 0.95 | 1.00 | 0.95 | 1.07 | *decr.* |
| 3 | 0.70 | 0.72 | 0.97 | 0.91 | *incr.* |
| 4 | 0.69 | 0.70 | 0.99 | 1.04 | *decr.* |
| 5 | 0.77 | 0.78 | 0.98 | 1.05 | *decr.* |
| 6 | 0.88 | 1.00 | 0.88 | 1.17 | *decr.* |
| 7 | 0.81 | 0.82 | 0.98 | 1.06 | *decr.* |
| 8 | 0.95 | 1.00 | 0.95 | 1.12 | *decr.* |
| 9 | 1.00 | 1.00 | 1.00 | 1.10 | *decr.* |
| 10 | 1.00 | 1.00 | 1.00 | 1.00 | *const.* |
| 11 | 0.98 | 0.99 | 1.00 | 0.95 | *incr.* |
| 12 | 1.00 | 1.00 | 1.00 | 1.00 | *const.* |
| 13 | 0.66 | 0.75 | 0.88 | 0.70 | *incr.* |
| 14 | 1.00 | 1.00 | 1.00 | 1.00 | *const.* |
| 15 | 0.77 | 0.78 | 0.99 | 0.92 | *incr.* |
| 16 | 0.94 | 0.96 | 0.98 | 0.88 | *incr.* |
| 17 | 1.00 | 1.00 | 1.00 | 1.00 | *const.* |
| 18 | 1.00 | 1.00 | 1.00 | 1.00 | *const.* |
| 19 | 0.90 | 0.93 | 0.97 | 0.87 | *incr.* |
| 20 | 0.68 | 0.79 | 0.87 | 0.69 | *incr.* |
| 21 | 0.98 | 1.00 | 0.98 | 0.92 | *incr.* |
| 22 | 1.00 | 1.00 | 1.00 | 1.00 | *const.* |
| 23 | 0.80 | 1.00 | 0.80 | 1.22 | *decr.* |
| 24 | 1.00 | 1.00 | 1.00 | 1.00 | *const.* |
| 25 | 1.00 | 1.00 | 1.00 | 1.00 | *const.* |
| 26 | 0.85 | 0.86 | 0.99 | 1.02 | *decr.* |
| 27 | 0.68 | 0.68 | 0.99 | 0.86 | *incr.* |
| 28 | 1.00 | 1.00 | 1.00 | 1.00 | *const.* |
| 29 | 0.65 | 0.69 | 0.94 | 0.75 | *incr.* |
| 30 | 0.90 | 0.93 | 0.98 | 0.90 | *incr.* |
| 31 | 1.00 | 1.00 | 1.00 | 1.00 | *const.* |
| 32 | 0.90 | 0.92 | 0.98 | 0.91 | *incr.* |
| **Average** | 0.88 | 0.92 | 0.96 | | |

**Table 3B**. Specific efficiency scores of Assistant professors

| Professor | $\theta^*_{CRS}$ | $\theta^*_{VRS}$ | SE | $\sum_{k=1}^{K} \lambda^*_k$ | Status |
|---|---|---|---|---|---|
| 1 | 0.94 | 0.99 | 0.95 | 1.11 | *decr.* |
| 2 | 0.92 | 0.92 | 0.99 | 0.93 | *incr.* |
| 3 | 0.76 | 0.79 | 0.97 | 0.84 | *incr.* |
| 4 | 0.84 | 0.91 | 0.93 | 0.82 | *incr.* |
| 5 | 0.70 | 0.81 | 0.86 | 0.60 | *incr.* |
| 6 | 1.00 | 1.00 | 1.00 | 1.00 | *const.* |
| 7 | 0.89 | 0.90 | 0.99 | 0.90 | *incr.* |
| 8 | 0.92 | 0.93 | 0.99 | 0.94 | *incr.* |
| 9 | 0.81 | 0.93 | 0.87 | 0.69 | *incr.* |
| 10 | 0.79 | 0.93 | 0.85 | 0.62 | *incr.* |
| 11 | 0.88 | 1.00 | 0.88 | 0.68 | *incr.* |
| 12 | 0.65 | 1.00 | 0.65 | 0.40 | *incr.* |
| 13 | 0.65 | 0.73 | 0.89 | 0.70 | *incr.* |
| 14 | 0.96 | 0.97 | 0.99 | 1.06 | *decr.* |
| 15 | 0.95 | 0.96 | 0.98 | 0.88 | *incr.* |
| 16 | 0.81 | 0.84 | 0.96 | 0.86 | *incr.* |
| 17 | 1.00 | 1.00 | 1.00 | 1.00 | *const.* |
| 18 | 1.00 | 1.00 | 1.00 | 1.00 | *const.* |
| 19 | 0.94 | 1.00 | 0.94 | 0.65 | *incr.* |
| 20 | 0.88 | 0.88 | 0.99 | 0.98 | *incr.* |
| 21 | 0.87 | 0.87 | 1.00 | 1.02 | *decr.* |
| 22 | 0.67 | 0.83 | 0.80 | 0.55 | *incr.* |
| 23 | 1.00 | 1.00 | 1.00 | 1.00 | *incr.* |
| 24 | 1.00 | 1.00 | 1.00 | 1.00 | *const.* |
| 25 | 0.76 | 0.80 | 0.96 | 0.78 | *incr.* |
| 26 | 1.00 | 1.00 | 1.00 | 1.00 | *const.* |
| 27 | 0.68 | 0.82 | 0.83 | 0.65 | *incr.* |
| 28 | 0.84 | 0.95 | 0.89 | 0.75 | *incr.* |
| 29 | 0.70 | 0.74 | 0.95 | 0.81 | *incr.* |
| 30 | 0.69 | 0.75 | 0.93 | 0.74 | *incr.* |
| 31 | 0.80 | 0.83 | 0.97 | 0.78 | *incr.* |
| **Average** | 0.85 | 0.91 | 0.94 | | |